# List of the close asteroid pairs strongly perturbed by three body resonances

[1]Rosaev A., [2]Galiazzo M., [3]Plavalova E

[1]Research and Educational Center "Nonlinear Dynamics", Yaroslavl State University
[2]Department of Astrophysics, University of Vienna, Austria
[3]Mathematical Institute Slovak Academy of Sciences

The two and three body mean motion resonances with Jupiter and Mars (and possible with Earth) may have significant effect on dynamics of very young close asteroids pairs and families. The most known example is influence 9:16 resonance with Mars to some members of Datura family. It is evident, that evolution of this family can be properly reconstructed only when this fact has taken into account. Here we give the list of asteroid pairs in vicinity of mean motion resonance.

One of the results of resonance perturbation may be the temporary compensation of the Yarkovsky drift in the semimajor axis. Evidently, studying this relationship may very well prove to be a radically influential factor in the age estimation of asteroid pairs and families. In this paper we have given list of some close asteroid pairs in vicinity of mean motion resonances.

### Introductions

Asteroids tend to groups into so-called families: the associations of objects sharing similar orbits. Most of them are the results of very old, about 1 Gyrs, collision between asteroids [1.2]. Since the beginning of the twenty century, asteroid families are the object of intensive studying.

A new and exciting development in the past decade was the detection of several asteroid families with very young formation ages. These cases are important, because various collisional and dynamical processes had little time to act on these families to alter their properties. Recent studies have shown that some asteroid families can also be the outcome of the spin-up-induced fission of a critically rotating parent body (fission clusters, [3]). Moreover, cases of subsequent breakup can take place in old families [4].

It is important, that the age of a young family can be determined by numerically integrating the orbits of its members backward in time and demonstrating that they converge to each other at some specific time in the past. The method of backward integration of orbits only works for the families with age smaller than few million years.

As it was noted in paper Broz and Vokrouhlicky [5], in resonance we have monotonic changes in eccentricity instead semimajor axis drift. As a result, resonances can occur significant effect on evolution some close asteroid pairs. It is expected, that mean motion resonances with Mars and three body resonances can provide the similar effect.

Some examples of the resonance perturbations of young families and pairs are known, first of al, it is Datura family in 9:16 resonance with Mars (Nesvorny, et al, 2006) [6]. Pravec et al. [7] note that pair (49791) 1999 XF31 and (436459) 2011 CL97 chaotic orbits may be explained by 15:8 mean motion resonance with Mars. Duddy et al. [8] pointed that pair (7343) Ockeghem and (154634) 2003 XX38 is in 2-1J-1M three body resonance.

The main goal of this paper is to search for the resonances, nearest to the known very young asteroid pairs. In result, we present the list of the asteroid pairs close to resonance and try to classify it by degree of perturbations.

### Methods and problem setting

To study the dynamic evolution of asteroid families in this paper, the equations of the motion of the systems were numerically integrated orbits over 800 kyr using the N-body integrator Mercury (Chambers 1999) [9] and the Everhart integration method [10].

To the nominal resonance position calculation, we use values of semimajor axis of planets, averaged over time of integration: 1.52368 AU for Mars, 5.20259 AU for Jupiter, 9.55490 AU for Saturn always in this paper. Where it is possible, we have compared our result with Smirnov and Schevchenko results [11] and note good agreement.

To study interaction considered pair with resonance and to determine position of resonance center (chaotic zone center) we apply integration of orbits of asteroid with significant values of Yarkovsky effect ($A_2 = 1*10^{-13}$).

**Results**

We test 18 pairs of asteroid from paper (Pravec et al, 2019) on subject neighboring to any two or three body mean motion resonance up to 20 order. Only one pair 21436 Chaoyichi -334916 (2003 YK39) is close to simple two body mean motion resonance 11:3 with Jupiter.

But in half of them we detect three body resonance in distance smaller than d=0.0006 AU at least for one asteroid of pair (table 1). These cases display strong resonance perturbations in semimajor axis (letter P in the last column of table). In some cases asteroids may display jumps from one side of resonance to another in dependence on Yarkovsky effect value. (letter J in the last column of table). Other studied pairs of asteroid are more distant from any resonances (d>0.0006 AU) and have not notable perturbations (table 2). Therefore, this value may be considered as a rough boundary since that resonance perturbations are insignificant.

Such large percent of the resonance-perturbed among the asteroid pairs allows us to suppose the relation of the resonances with the process of pairs forming. In the light of this relation, the new theoretical researches of the process of the origin of asteroid pairs are strongly required.

On the other hand, compact asteroid families and pairs near resonance give the unique possibility to study in detail resonance perturbations and dynamical interaction of minor body with resonance.

As it is possible to see in the table, only in five cases (about 25%) we have Jupiter-Saturn-asteroid resonance. All other – resonances with Earth group planets: (Earth or Mars) – Jupiter- asteroid. It highlights the role of Earth group planets on the dynamics in the inner asteroid belt.

Here we have shown that at least in case of pair (44620) 1999 RS43 and (295745) 2008 UH98 resonance can dampen Yarkovsky effect and prevent Yarkovsky related semimajor axis drift.

**Pair (44620) 1999 RS43 and (295745) 2008 UH98**

This pair was studied by Pravec et al (2019). They found that the primary (44620) 1999 RS43 is a binary system. The synchronous satellite (bound secondary) has the secondary-to-primary mean diameter ratio D1,s/D1,p = 0.39, an orbital period of 33.64 h and a prograde orbit. The primary's rotational period is about P1,p = 3.14 h. Pravec et al estimate epoch of origin of pair about 700 kyr ago.

Currently, the semimajor axis of the (44620) 1999 RS43 is slightly smaller than that of the (295745) 2008 UH98 and this difference is of the order of $3*10^{-4}$ AU. At the same time, the diameter of (44620) 1999 RS43 is significantly larger than the diameter of (295745) 2008 UH98. Accordingly, the expected value of the semimajor axis drift due to the Yarkovsky effect in (295745) 2008 UH98 is greater. Therefore, we can exclude all values of $da/dt_{295745} < 0$ (A2<0) since, herewith, the orbits in the pair diverge.

At the numerical integrations, we use non-gravitational parameter $A_2$. In general, at fixed eccentricity, the rate of Yarkovsky drift da/dt is proportional to $A_2$. The dependence between da/dt and $A_2$ has built on our data of numeric integration orbit (295745) 2008 UH98 is given in fig.1 For the secondary (295745) 2008 UH98 the dependence between $A_2$ and d$a$/dt can be linearly approximated: d$a$/dt =9771*( $A_2$)+1362 (Fig.1). But in the studied pair only smaller asteroid shows simple linear Yarkovsky drift: for primary (44620) 1999 RS43 linear approximation of this dependence is not possible. The motion of (44620) 1999 RS43 is strongly chaotic for different values of coefficient $A_2$ (Fig.2). We explain this fact by some resonance effect.

However, in our search for simple two body mean motion resonance, we obtain for closest resonance very high order (85:23J) which is unreal. By this reason we continue our search. To do it, we repeat our integrations with different perturbations: only one large planet; two planets. Only in case perturbation (44620) 1999 RS43 orbit by Jupiter and Saturn we obtain resonance perturbations (Fig.3). However, it is still high order resonance not included in Gallardo atlas [12]. One candidate is 1J-7S-2 but it nominal position (2.174903 AU) is relatively far from (44620) 1999 RS43 orbit. But, definitely, it is three body Jupiter-Saturn-asteroid resonance: most probable it is 4S-9J-2 at a=2.176198 AU.

Table 1. Results of three-body mean motion resonance search

| # | Asteroid pair | Proper $a$, AU | Resonance | Nominal resonance position, AU | | Age, kyr | Comment |
|---|---|---|---|---|---|---|---|
| | | | | Present paper | [11] | | |
| 1 | (44620) 1999 RS43 | 2.17644 | 2+9J-4S | 2.176198 | - | 700 | P |
| | (295745) 2008 UH98 | 2.17669 | | | | | |
| 2 | (7343) Ockeghem | 2.19254 | 2-1J-1M | 2.192728 | 2.192762 | >382 | J |
| | (154634) 2003 XX38 | 2.19253 | | | | | |
| 3 | 6369 (1983 UC) | 2.29324 | 7-5J-3M | 2.292687 | 2.292729 | 700 | J |
| | 510132 (2010 UY57) | 2.29315 | | | | | |
| 4* | 404118 (2013 AF40) | 2.21744 | 2+10J-7S | 2.217453 | - | ? | J |
| | 355258 (2007 LY4) | 2.21746 | | | | | |
| 5 | 4765 Wasserburg | 1.94542 | 5-10J-1E | 1.945479 | - | 200 | P |
| | 350716 (2001 XO105) | 1.94563 | | | | | |
| 6 | 8306 Shoko | 2.24159 | 3-9J -4S | 2.241658 | - | 400 | J |
| | 2011 SR158 | 2.24125 | | | | | |
| 7 | 80218 (1999 VO123) | 2.2185 | 1+6J-6S | 2.219288 | 2.221212 | 140 | J |
| | 213471 (2002 ES90) | 2.21864 | | | | | |
| 8 | (43008) 1999 UD31 | 2.3481 | 2+7J -1S | 2.347592 | 2.347954 | 270 | P |
| | (441549) 2008 TM68 | 2.34773 | | | | | |
| 9 | (26420) 1999 XL103 | 2.19757 | 6-10J – 1E | 2.196919 | - | 250 | P |
| | 2012 TS209 | 2.19749 | | | | | |
| 10 | 49791 (1999 XF31) | 2.31665 | 8- 8J-3M | 2.316370 | - | 200 | P |
| | 436459 (2011 CL97) | 2.31663 | | | | | |

Table 2. Asteroid pairs have no any resonance perturbations

| # | Asteroid pair | Proper $a$, AU | Resonance | Nominal resonance position (Rosaev), AU | Proposed age of pair, Kyr |
|---|---|---|---|---|---|
| 11 | 25021 Nischaykumar | 2.31788 | 9-5J-4M | 2.319260 | 900 |
| | 453818 (2011 SJ109) | 2.31779 | | | |
| 12 | 26416 (1999 XM84) | 2.34257 | 6- 8J-1E | 2.341984 | 270 |
| | 214954 (2007 WO58) | 2.34256 | | | |
| 13 | 4905 Hiromi | 2.60102 | 4- 5J-1M | 2.601956 | 1800 |
| | 7813 Anderserikson | 2.60112 | | | |
| 14 | 2110 Moore-Sitterly | 2.19804 | 6-10J-1E | 2.196919 | 2000 |
| | 44612(1999 RP27) | 2.19787 | | | |
| 15 | 46829 McMahon | 2.39991 | 9-5J-2E | 2.399548 | 800 |
| | 2014 VR4 | 2.40048 | | | |
| 16 | 42946 (1999 TU95) | 2.56782 | 9- 7J- 3M | 2.569604 | 740 |
| | 165548 (2001 DO37) | 2.56761 | | | |
| 17 | 17198 Gorjup | 2.27969 | 10-3J-5M | 2.276560 | 300 |
| | 229056 (2004 FC126) | 2.27962 | | | |

This example is very important because demonstrate compensation Yarkovsky effect (at least in finite time interval) by resonance. Moreover, it shows that high order mean motion resonances can act very locally, on the one member of close pair.

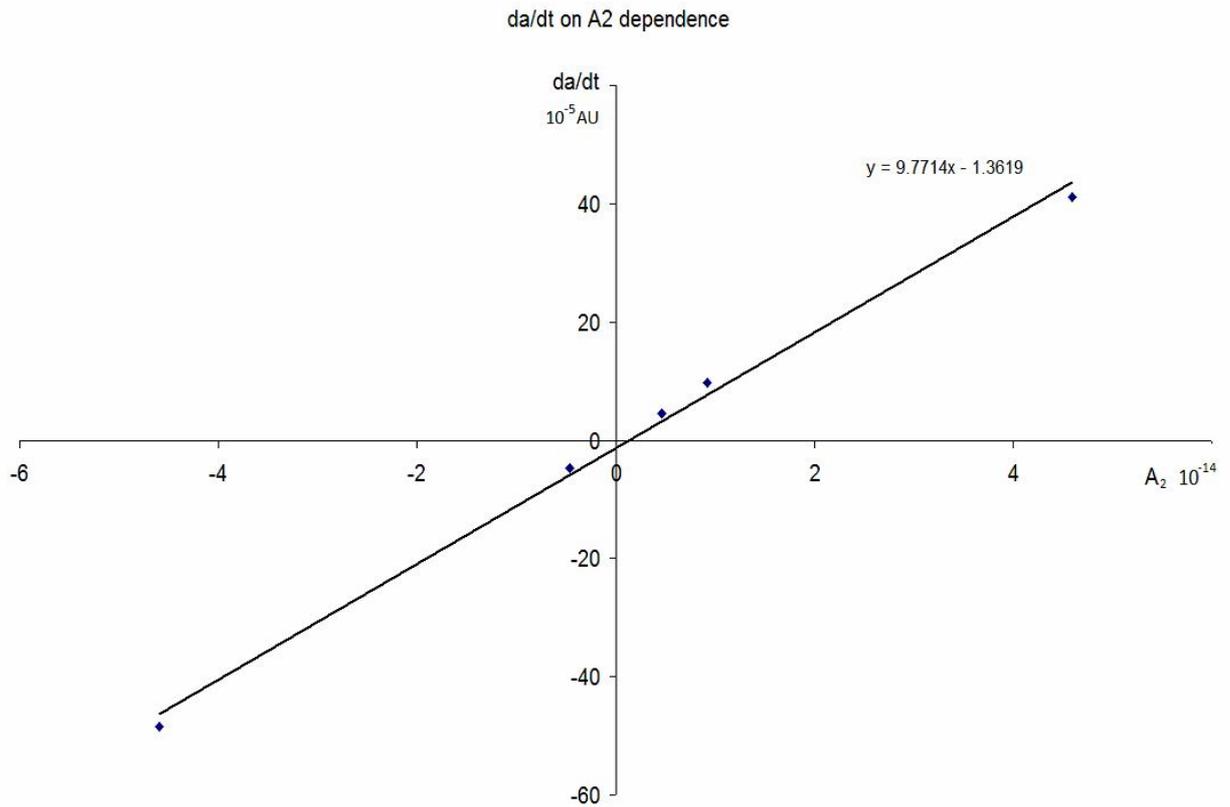

Fig.1. da/dt on $A_2$ dependence for **(295745) 2008 UH98**

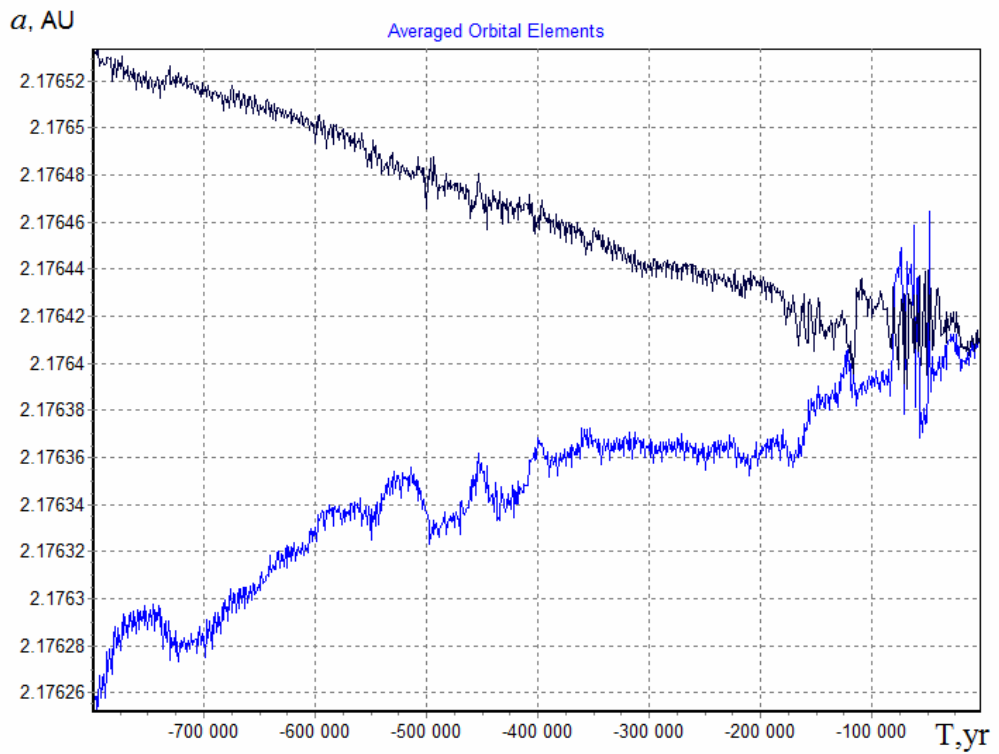

Fig.2. **(44620) 1999 RS43** Yarkovsky drift in semimajor axis $A_2=\pm1.8*10^{-14}$. All planets perturbations

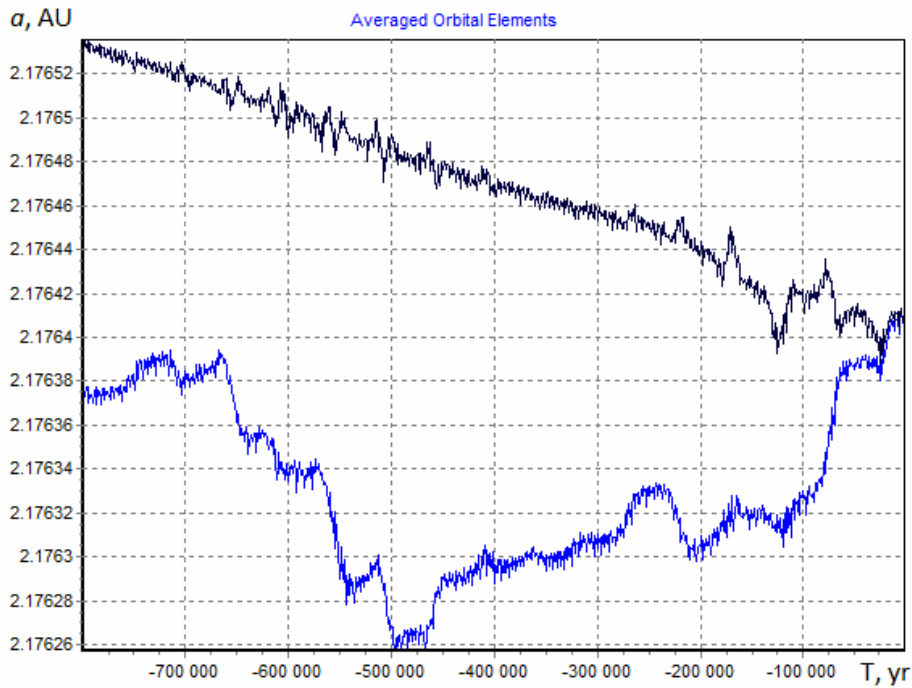

Fig.3. **(44620) 1999 RS43** Yarkovsky drift in semimajor axis $A_2=\pm1.8*10^{-14}$. Saturn and Jupiter perturbations

**Conclusions**

In this paper we have given list of some close asteroid pairs in vicinity of mean motion resonances. We have obtained that resonances affect on close asteroid pairs in a distance smaller than 0.0006 AU. 2) Only one case of important resonances is simple two body mean motion resonance, all other – three body. 3) Earth-Jupiter- asteroid and Mars-Jupiter-asteroid resonances are very important in the close asteroid pair dynamics in the inner and middle asteroid belt. In some cases perturbing resonances may be very high order.

We conclude that resonance perturbations may play an important role in dynamical evolution of very young asteroid families and close pairs. It is necessary to account this fact in the models of their origin and dynamical evolutions.